\documentclass{article}

\usepackage{graphicx}
\usepackage{amsmath}
\usepackage{amssymb}
\usepackage{tensor}
\usepackage{url}
\usepackage{authblk}

\newcommand{\sd}{\mathrm{d}}

\newcommand{\bra}[1]{\left<#1\right|}
\newcommand{\ket}[1]{\left|#1\right>}

\title{Accelerating decay with acceleration}
\author[1]{Wim Beenakker}
\author[1]{David Venhoek}
\affil[1]{Theoretical High Energy Physics, Radboud University
Nijmegen, Heyendaalseweg 135, 6525~AJ Nijmegen, The Netherlands}
\date{10 October 2023}

\allowdisplaybreaks

\begin{document}

\maketitle

\begin{abstract}
We investigate accelerated Unruh-deWitt detectors as a model for particle decay. We find non-trivial decay rates, including a pattern of peaks in decay rate that extends to lower accelerations. Applying our model to the alpha decay of $\mathrm{^{210}Po}$, we find that effects could be observed with an acceleration of $a\approx 10^{26} \frac{\mathrm{m}}{\mathrm{s}^2}$ as long as that acceleration is controlled to within 1 percent. Although still out of reach of current experimental setups, other decay processes at lower energy, such as beta decay, could result in the peaks we find being within range of future experiments.
\end{abstract}

The Unruh effect, discovered in the 70s by Fulling \cite{Fulling} and Unruh \cite{Unruh}, is the prediction that an accelerating observer will, in the Minkowski vacuum, see particles as a consequence of being accelerated. As the Unruh effect has significant similarities to predictions such as Hawking radiation, it has drawn considerable attention over the years, a relatively recent overview of which can be found in \cite{Crispino}.

There have been investigations into methods of observing the Unruh effect as early as the 80s by Bell et al. \cite{bell1,bell2}, but direct observation seems to require very large acceleration, $a \approx 10^{30} \frac{\mathrm{m}}{\mathrm{s}^2}$, and has some problems with thermal effects giving a similar signature. There have, however, been some claims of an observation of the Unruh effect in data from the NA63 experiment at CERN \cite{Lynch1, Lynch2}.

Recently, suggestions have been made for more indirect methods of observing the Unruh effect by investigating the effects of acceleration on entanglement of multiple detectors \cite{Zhou}. The situation of an accelerated detector in a thermal bath has also received attention \cite{Garay}, giving rise to cooling in certain parameter regions. Such indirect observation approaches, whilst not showing directly the original predictions, still rely on the same underlying physics, and provide alternate pathways for probing it.

Decay of unstable particles and nuclei could provide another such pathway. The effects of acceleration on decay have been previously studied in \cite{Muller}, which showed deviations when the acceleration is of the same order as the mass of the decaying particle, although its approach required ignoring the mass of all but the heaviest decay product. This approach was expanded upon in \cite{Matsas}, although its interpretation is mudied by the use of a 4-dimensional fermion model in a 2-dimensional spacetime.

In this paper, we will look in more detail at Unruh-deWitt detectors \cite{Unruh,DeWitt} as a model for decay, attempting to get another approach for experimental observation of effects related to the Unruh effect. In contrast to the approach in \cite{Muller}, the use of Unruh-deWitt detectors allows us to consider a second decay product with mass, at the cost of assuming the heaviest decay product's trajectory is not significantly modified. We will then use this model to make predictions on alpha decay, specifically of $\mathrm{^{210}Po}$.

In Sections~\ref{sec:model} through~\ref{sec:accelerated}, we focus on setting up our model and working out its consequences in both the accelerating and non-accelerating case. For ease of calculation this will be done in natural units. In Section~\ref{sec:phys} we then reintroduce all relevant constants, and use our model to investigate the effects of acceleration on the decay of $\mathrm{^{210}Po}$, and what accelerations are needed to measure these effects. We conclude by discussing the implications of our results in Section~\ref{sec:conclusion}, discussing potential roads to observing the found effect and the open theoretical questions still present.

\section{Detector model}\label{sec:model}

For the rest of this paper, we will work with a massive real scalar field with mass $m$ in a 1+1 dimensional Minkowski space, coupled to a two-level Unruh-deWitt detector with energy gap $\Delta$, denoting its excited state with $\ket{1}$ and its ground state with $\ket{0}$. We assume this detector moves along an eigen-time parameterized, timelike path $x(\tau)$. Further, we take the detector to couple to the field through a time independent dimensionless operator $M$, which while we keep it general here, will be of the form $\begin{pmatrix}0 & 1\\1 & 0\end{pmatrix}$ in many applications. This leaves a coupling strength $G$ with mass-dimension $1$. Combined, this gives the action
\begin{align}
S &= \int\sd^2 x \left(\frac{1}{2}\eta\indices{^\mu^\nu}\partial\indices{_\mu}\phi\partial\indices{_\nu}\phi-\frac{1}{2}m^2\phi^2\right) + \int\sd \tau \left(\Delta\ket{1}\bra{1} + GM\phi(x(\tau))\right).
\end{align}

Moving to the interaction picture, time evolution of a state $\ket{s}$ is then given by
\begin{align}
\frac{\sd}{\sd \tau}\ket{s} &= -iGM(\tau)\phi(x(\tau))\ket{s}.
\end{align}
where
\begin{align}
M\left(\tau\right) &= e^{i\Delta\tau\ket{1}\bra{1}}Me^{-i\Delta\tau\ket{1}\bra{1}}
\end{align}

Using this, we can calculate the probability of a detector in the excited state, moving through the Minkowski vacuum, decaying to its ground state after some (eigen)time $T$\footnote{Throughout this paper we will use heaviside switching for simplicity, as our main results don't significantly change for smoother switching functions.}. To lowest non-vanishing order in $G$, this is given by
\begin{align}
P_{10}(T) &= \sum_{\bra{\phi}}\left|\bra{0}\otimes\bra{\phi}\int_0^T\sd\tau GM(\tau)\phi(x(\tau))\ket{1}\otimes\ket{\phi_0}\right|^2\\
&= G^2\left|\bra{0}M(0)\ket{1}\right|^2\int_0^T\sd\tau_1\int_0^T\sd\tau_2e^{i\Delta(\tau_1-\tau_2)}\bra{\phi_0}\phi(x(\tau_1))\phi(x(\tau_2))\ket{\phi_0}
\end{align}
where $\ket{\phi_0}$ is the Minkowski vacuum state, and the sum is over all field states.

Since we are working in flat spacetime, and since $x(\tau)$ is a timelike path, Lorentz symmetry implies that $\bra{\phi_0}\phi(x(\tau_1))\phi(x(\tau_2))\ket{\phi_0}$ only depends on the (timelike) distance between $x(\tau_1)$ and $x(\tau_2)$. Furthermore, we will restrict to paths $x(\tau)$ of constant acceleration, which implies that the distance between $x(\tau_1)$ and $x(\tau_2)$ only depends on the difference $\tau_2-\tau_1\equiv\kappa$. Together, this simplifies the expression of the probability further to
\begin{align}
P_{10}(T) &= G^2\left|\bra{0}M(0)\ket{1}\right|^2\int_{-T}^T\sd\kappa \left(T-\left|\kappa\right|\right)e^{-i\Delta\kappa}g(\Delta_\tau(\kappa))
\end{align}
where
\begin{align}
g(\Delta_\tau) &= \frac{1}{2\pi}\int_0^\infty\sd p \frac{e^{i\sqrt{p^2+m^2}\Delta_\tau}}{\sqrt{p^2+m^2}}
\end{align}
and $\Delta_\tau(\kappa) = \kappa$ for a non-accelerated detector, and $\Delta_\tau(\kappa) = \frac{2}{a}\sinh\left(\frac{a\kappa}{2}\right)$ when accelerating with constant acceleration $a$.

Using integral representations of the Bessel functions (\cite{DLMF} 10.9.12) and the fact that $\Delta_\tau(-\kappa) = -\Delta_\tau(\kappa)$ we can rewrite this as
\begin{align}
\Gamma\left(a,T\right) = \frac{P_{10}(T)}{G^2\left|\bra{0}M(0)\ket{1}\right|^2T} &= \frac{1}{2T}\int_0^T\sd\kappa \left(T-\kappa\right)\sin\left(\Delta\kappa\right)J_0\left(m\Delta_\tau\left(\kappa\right)\right)\nonumber\\
&\phantom{=}-\frac{1}{2T}\int_0^T\sd\kappa\left(T-\kappa\right)\cos\left(\Delta\kappa\right)Y_0\left(m\Delta_\tau\left(\kappa\right)\right)\label{eq:decay},
\end{align}
which is the form which we will use throughout the rest of this paper.

\section{Stationary detector}\label{sec:unaccelerated}

First, let us consider the case of a stationary detector, which through use of Lorentz symmetry is representative of all non-accelerating detectors. The relationship between eigentime and timelike distance travelled then simplifies to $\Delta_\tau(\kappa) = \kappa$, and our main equation becomes
\begin{align}
\Gamma(0, T) &= \frac{1}{2T}\int_0^T\sd\kappa \left(T-\kappa\right)\sin\left(\Delta\kappa\right)J_0\left(m\kappa\right)\nonumber\\
&\phantom{=}-\frac{1}{2T}\int_0^T\sd\kappa\left(T-\kappa\right)\cos\left(\Delta\kappa\right)Y_0\left(m\kappa\right)\label{eq:prob_nonaccel}
\end{align}
which, taking the limit $T\rightarrow\infty$ (see Appendix~\ref{app:inftimelimit}), yields
\begin{align}
\lim_{T\rightarrow\infty} \Gamma(0,T) &= \begin{cases}
0 & \text{if } m > \Delta\\
\frac{1}{\sqrt{\Delta^2-m^2}} & \text{if } m < \Delta
\end{cases}
\end{align}

\begin{figure}[htp]
\center{\includegraphics[width=0.9\textwidth]{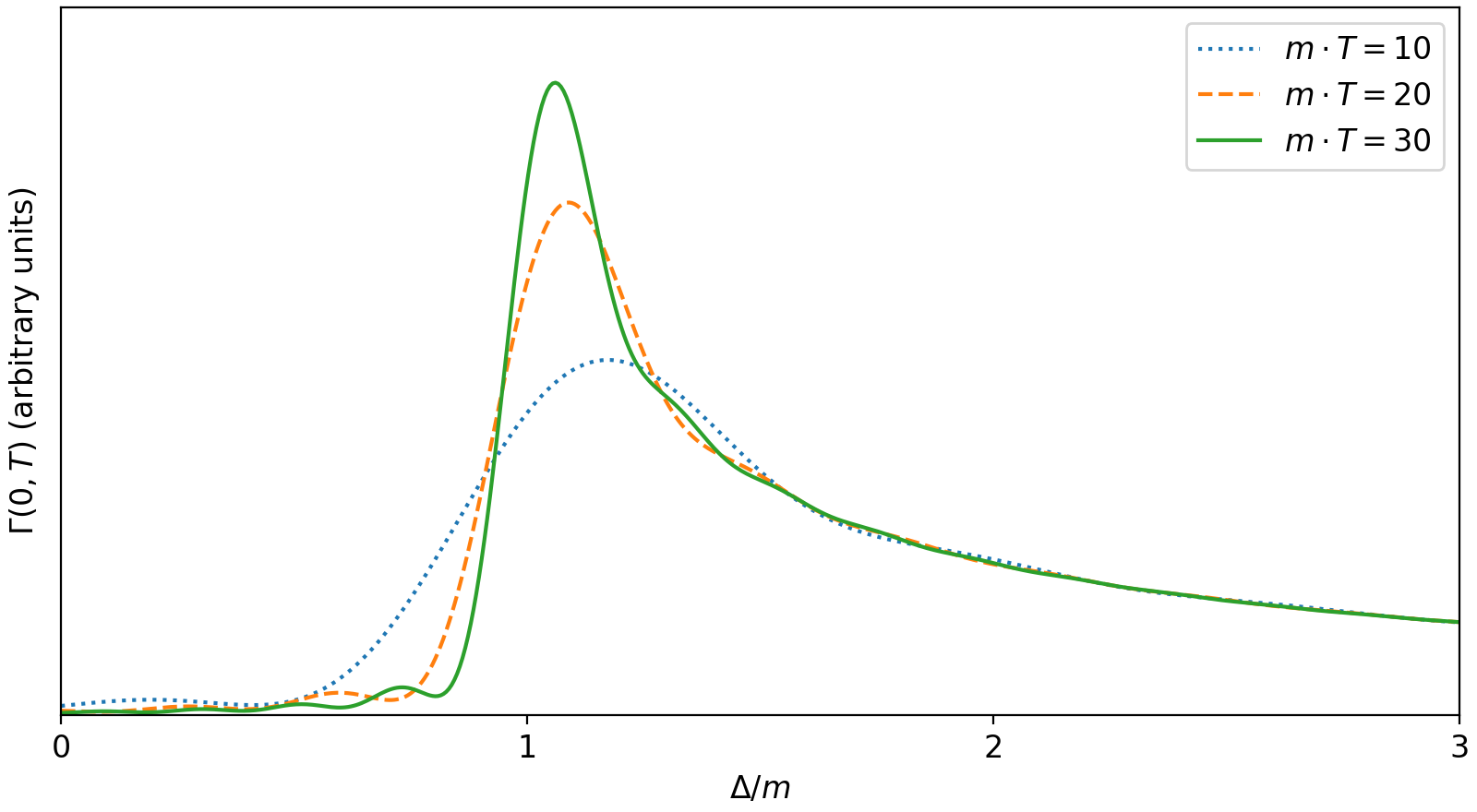}}
\caption{Decay rates for stationary detectors with varying detector energy gap, for an interaction lasting for a dimensionless time span $m\cdot T=10$, $20$ and $30$.}\label{fig:unaccelerated-fin}
\end{figure}

\begin{figure}[htp]
\center{\includegraphics[width=0.9\textwidth]{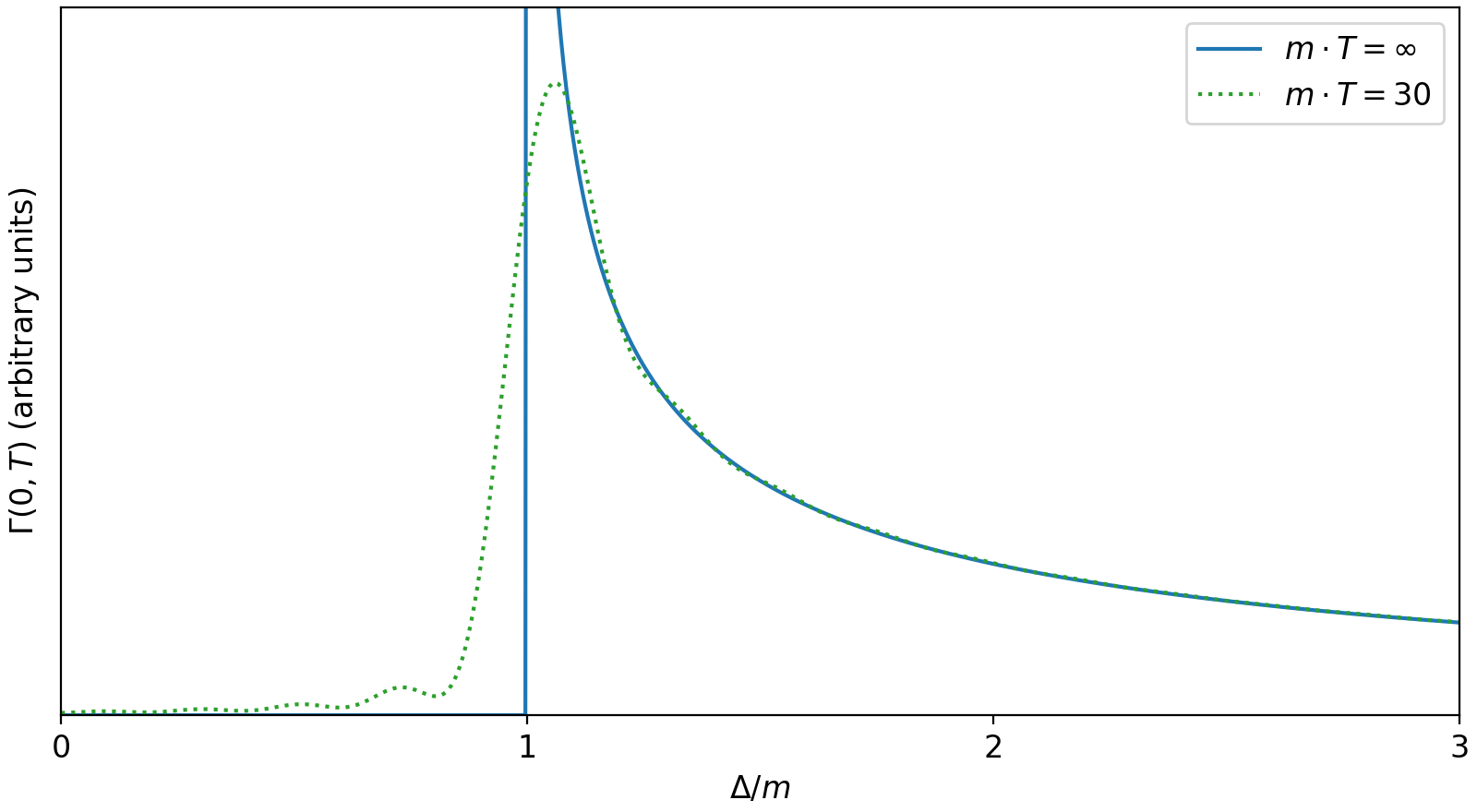}}
\caption{Infinite-time limit of the decay rate of a stationary detector with varying detector energy gap. For comparison, the result for an interaction lasting $m\cdot T=30$ is also shown.}\label{fig:unaccelerated-lim}
\end{figure}

Let us first consider how the decay rate changes as a function of the detector energy gap, plotted in Figure~\ref{fig:unaccelerated-fin}. We see a cutoff when this energy is the same as the mass of the particle. Below this cutoff decay is surpressed, as the detector does not contain sufficient energy to create particles in the field. When $\Delta \approx m$, the detector can resonate with the field, greatly enhancing the decay rate, which then drops off as the energy gap increases. An interesting feature here is that the width of the resonance decreases as $T$ increases, forming an illustration of energy-time uncertainty. On top of this global behaviour, we see smaller oscillations in the decay rate, that decrease in amplitude as $T$ increases.

This analytic expression of the $T\rightarrow\infty$ limit, plotted in Figure~\ref{fig:unaccelerated-lim}, clearly shows the cutoff when $\Delta < m$, the resonance at $\Delta=m$ and a one over momentum decay beyond that point. The latter becomes intuitive when realising that, since only one particle is produced, any momentum of that particle needs to be extracted from the force holding the detector in place. This extraction inhibits decay as the required amount of momentum increases.

\section{Accelerated detector}\label{sec:accelerated}

We now move on to a detector accelerated with constant acceleration $a$. This has the more complicated eigentime-distance relation $\Delta_\tau(\kappa) = \frac{2}{a}\sinh\left(\frac{a\kappa}{2}\right)$. The decay rate is then given by
\begin{align}
\Gamma(a, T) &= \frac{1}{2T}\int_0^T\sd\kappa \left(T-\kappa\right)\sin\left(\Delta\kappa\right)J_0\left(\frac{2m}{a}\sinh\left(\frac{a\kappa}{2}\right)\right)\nonumber\\
&\phantom{=}-\frac{1}{2T}\int_0^T\sd\kappa\left(T-\kappa\right)\cos\left(\Delta\kappa\right)Y_0\left(\frac{2m}{a}\sinh\left(\frac{a\kappa}{2}\right)\right)\label{eq:prob_accel}.
\end{align}

Taking the $T\rightarrow\infty$ limit, this becomes
\begin{align}
\lim_{T\rightarrow\infty}\Gamma(a,T) &= \frac{1}{a\pi} e^{\frac{\pi\Delta}{a}} \left(K_{\frac{i\Delta}{a}}\left(\frac{m}{a}\right)\right)^2
\end{align}
as worked out in Appendix~\ref{app:inftimelimit}, where $K$ is the modified bessel function of the second kind.

\begin{figure}[htp]
\center{\includegraphics[width=0.9\textwidth]{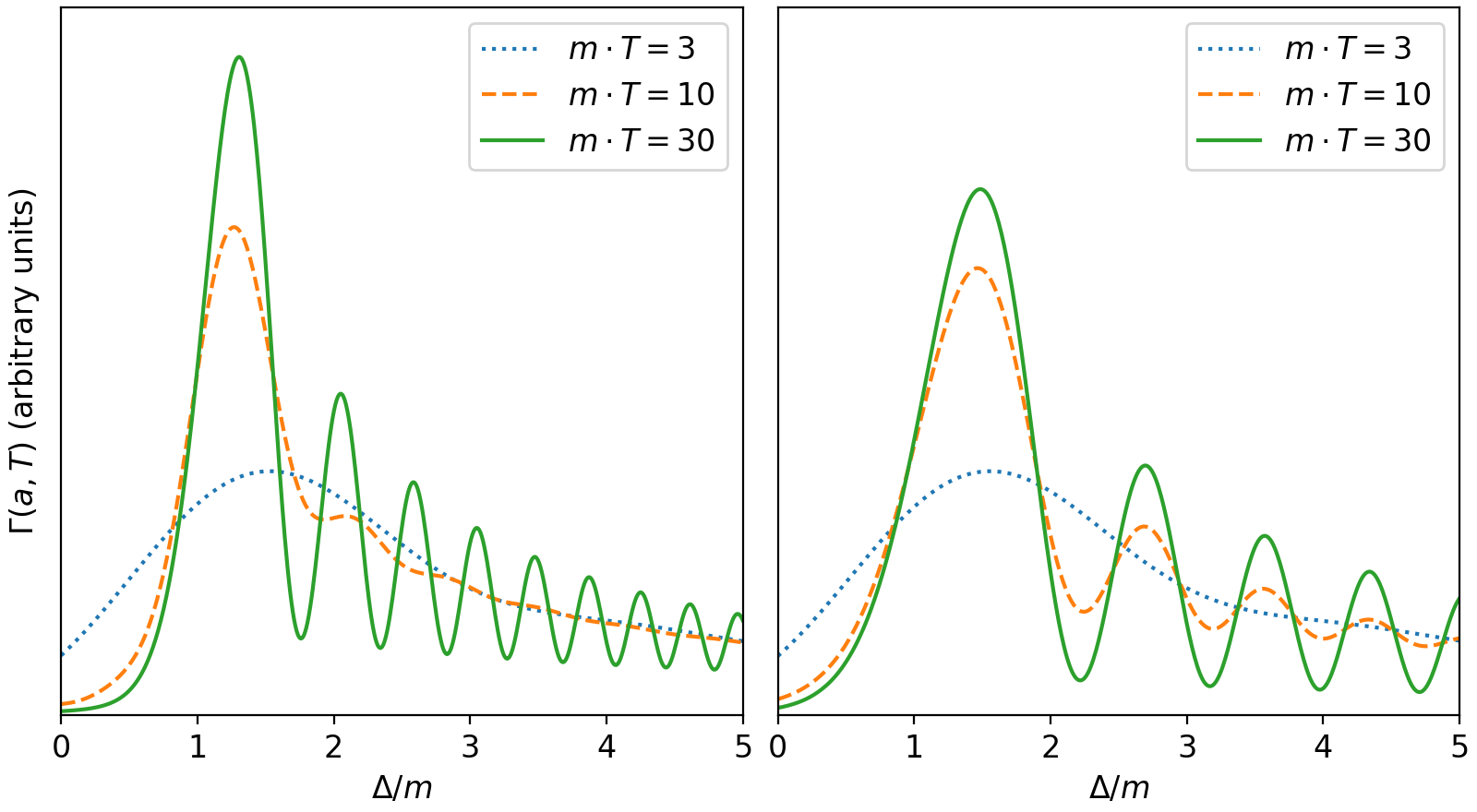}}
\caption{Decay rate of an accelerated detector with varying detector energy gap. On the left, acceleration $\frac{a}{m}=0.25$, on the right $\frac{a}{m}=0.5$.}\label{fig:accelerated_delta}
\end{figure}

\begin{figure}[htp]
\center{\includegraphics[width=0.9\textwidth]{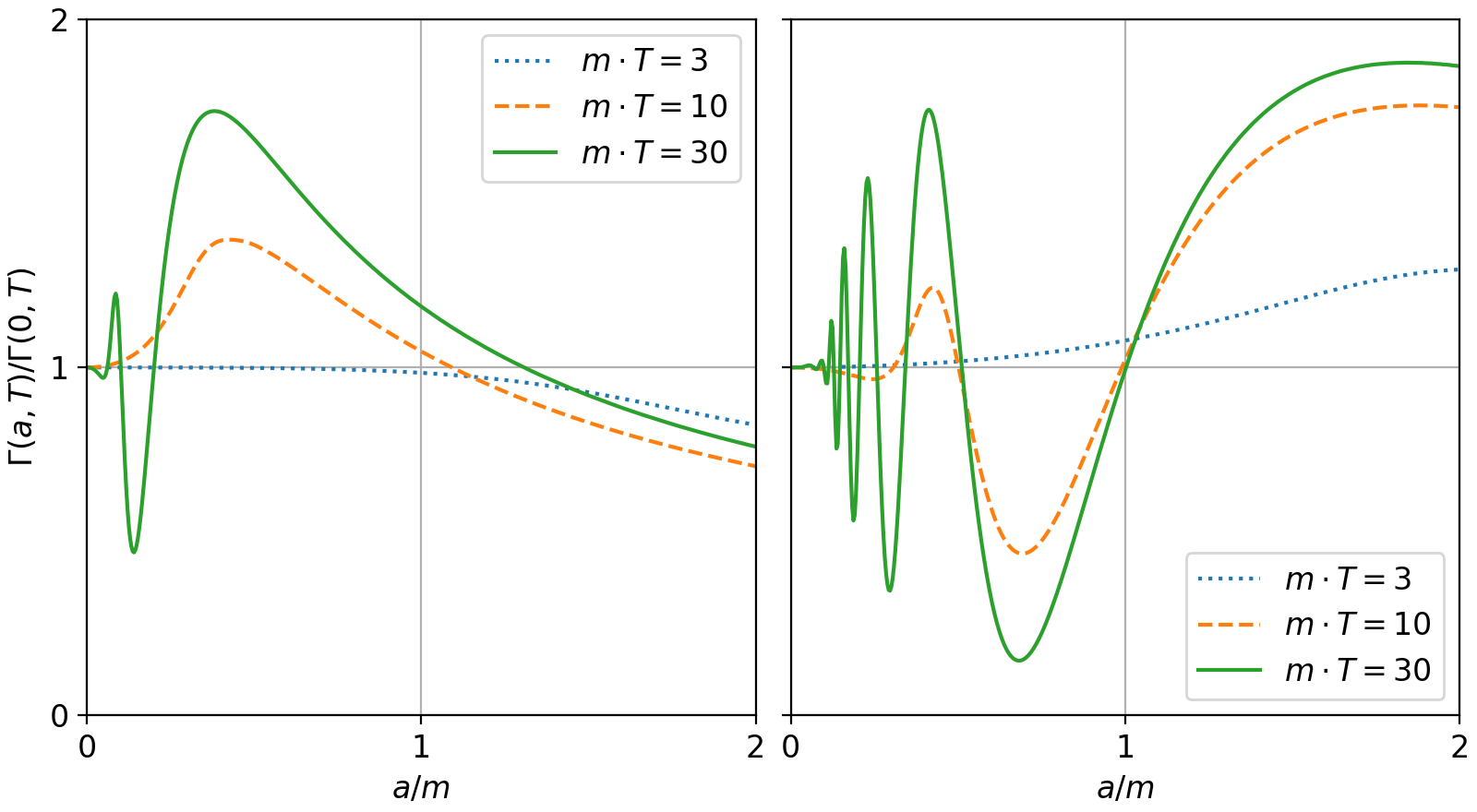}}
\caption{Decay rate, as a function of acceleration, relative to the unaccelerated decay rate. On the left $\frac{\Delta}{m} = 1.5$, on the right $\frac{\Delta}{m} = 2.5$.}
\label{fig:accelerated_a}
\end{figure}

Figure~\ref{fig:accelerated_delta} again shows the behaviour of the detector as a function of the detector energy gap. Note that the behaviour is markedly different to that seen in the stationary detector in Section~\ref{sec:unaccelerated}. Rather than a single resonance peak around the mass of the particle, we now see multiple resonance peaks. Furthermore, the spacing of these peaks depends on the acceleration experienced by the detector, becoming larger with increasing acceleration.

Note that compared with the unaccelerated result, finite-time effects here look quite different. In the unaccelerated case, we see on top of the developing peak and falloff a small $\Delta$-dependent wiggle. Here, we in contrast see no such additional wiggle, and we just see that the resonance peaks need time to develop. Furthermore, this development takes more time for the later peaks. That this is not just a result from being at a higher energy can be seen in Figure~\ref{fig:accelerated_a}, where the acceleration dependence of the decay rate is plotted for a given decay energy.

\begin{figure}[htp]
\center{\includegraphics[width=0.9\textwidth]{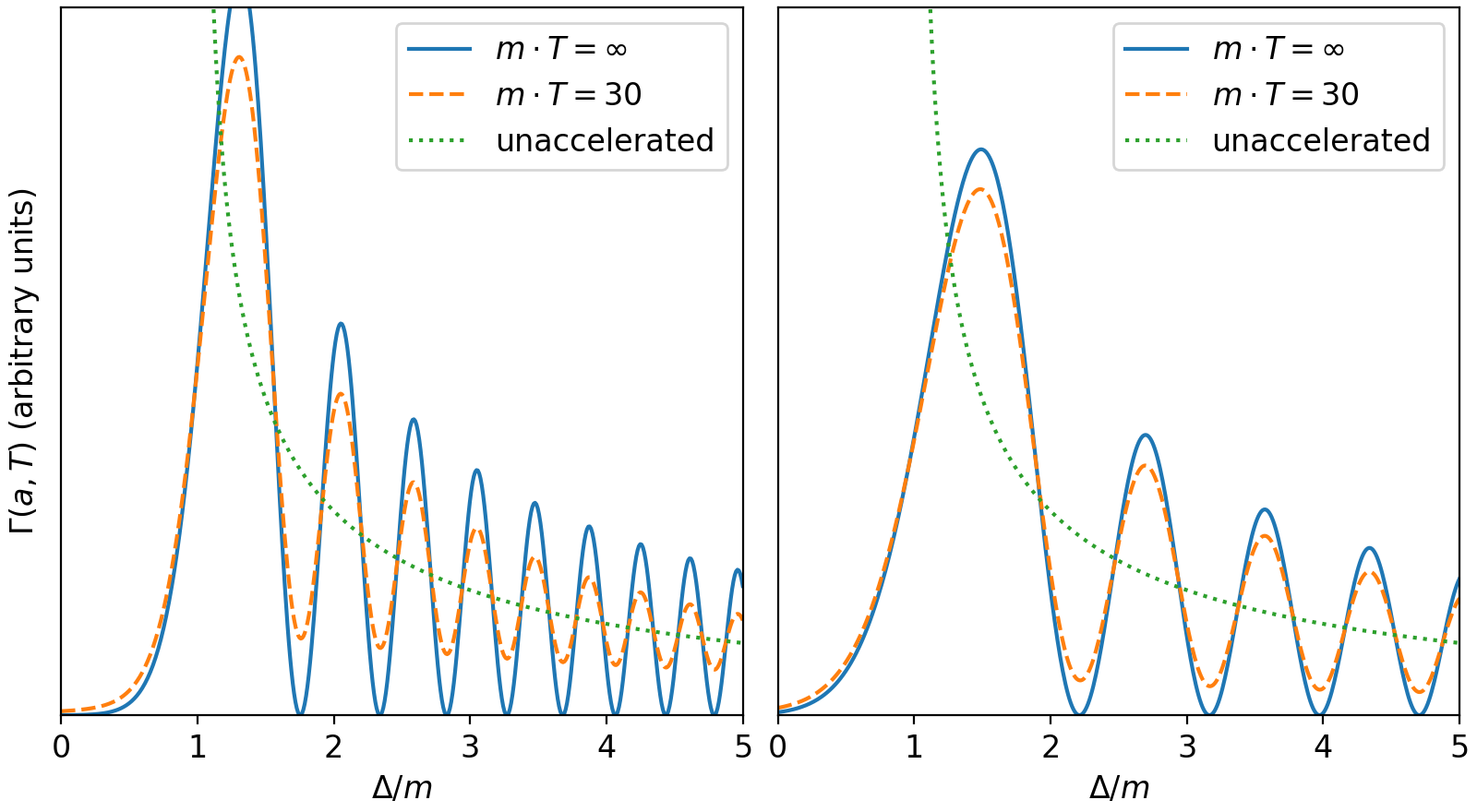}}
\caption{Infinite-time limit of the decay rate of an accelerated detector with varying detector energy gap. On the left acceleration $\frac{a}{m}=0.25$, on the right $\frac{a}{m}=0.5$. The dotted line gives the decay for an unaccelerated detector with the same mass and energy gap.}
\label{fig:accelerated_infdelta}
\end{figure}
\begin{figure}[htp]
\center{\includegraphics[width=0.9\textwidth]{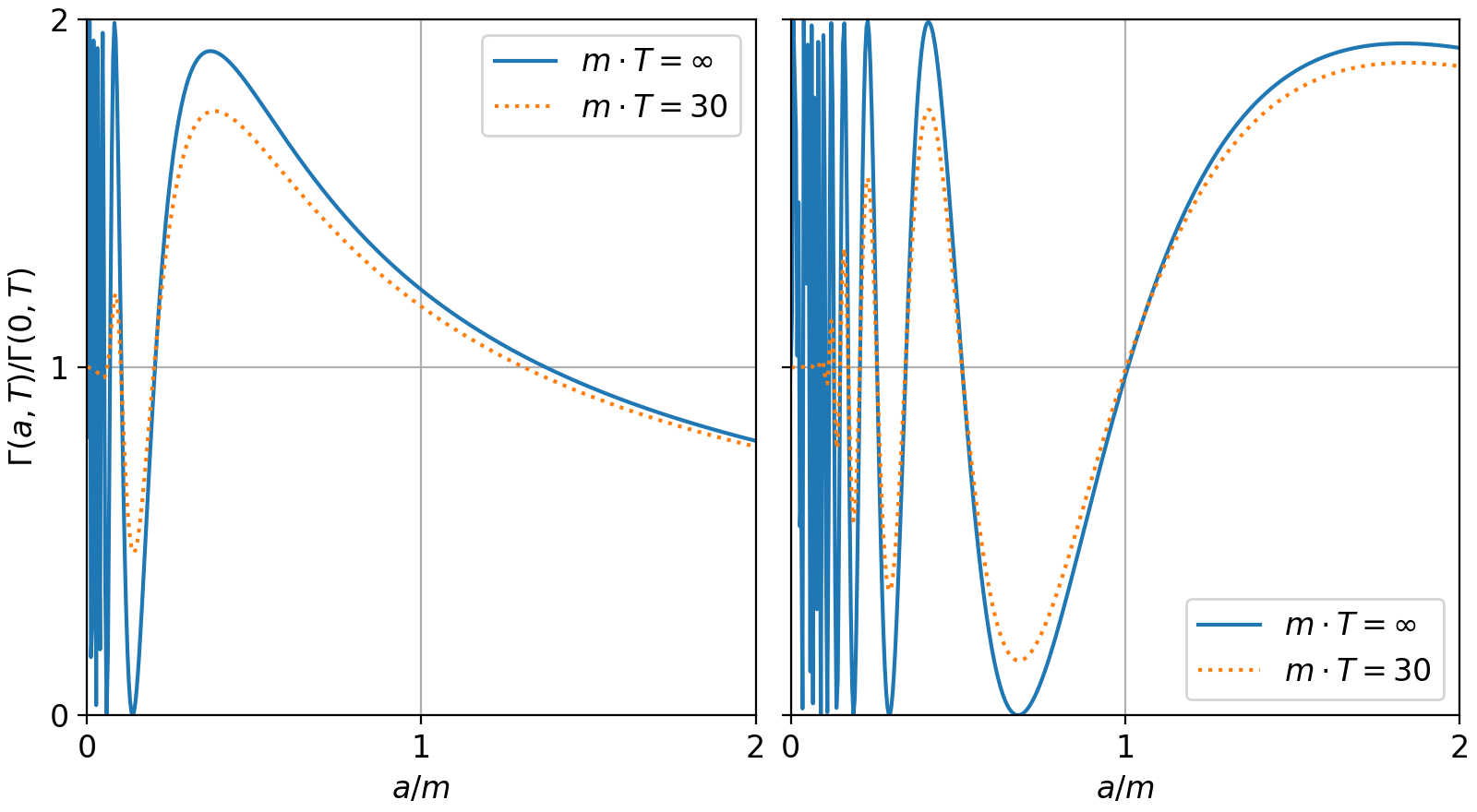}}
\caption{Infinite-time limit of the decay rate of an accelerated detector as function of acceleration, relative to the unaccelerated decay. On the left $\frac{\Delta}{m}=1.5$, on the right $\frac{\Delta}{m}=2.5$. Note that the rapid oscillations as $a$ goes to $0$ are not fully captured due to sampling limits.}
\label{fig:accelerated_infa}
\end{figure}
\begin{figure}[htp]
\center{\includegraphics[width=0.9\textwidth]{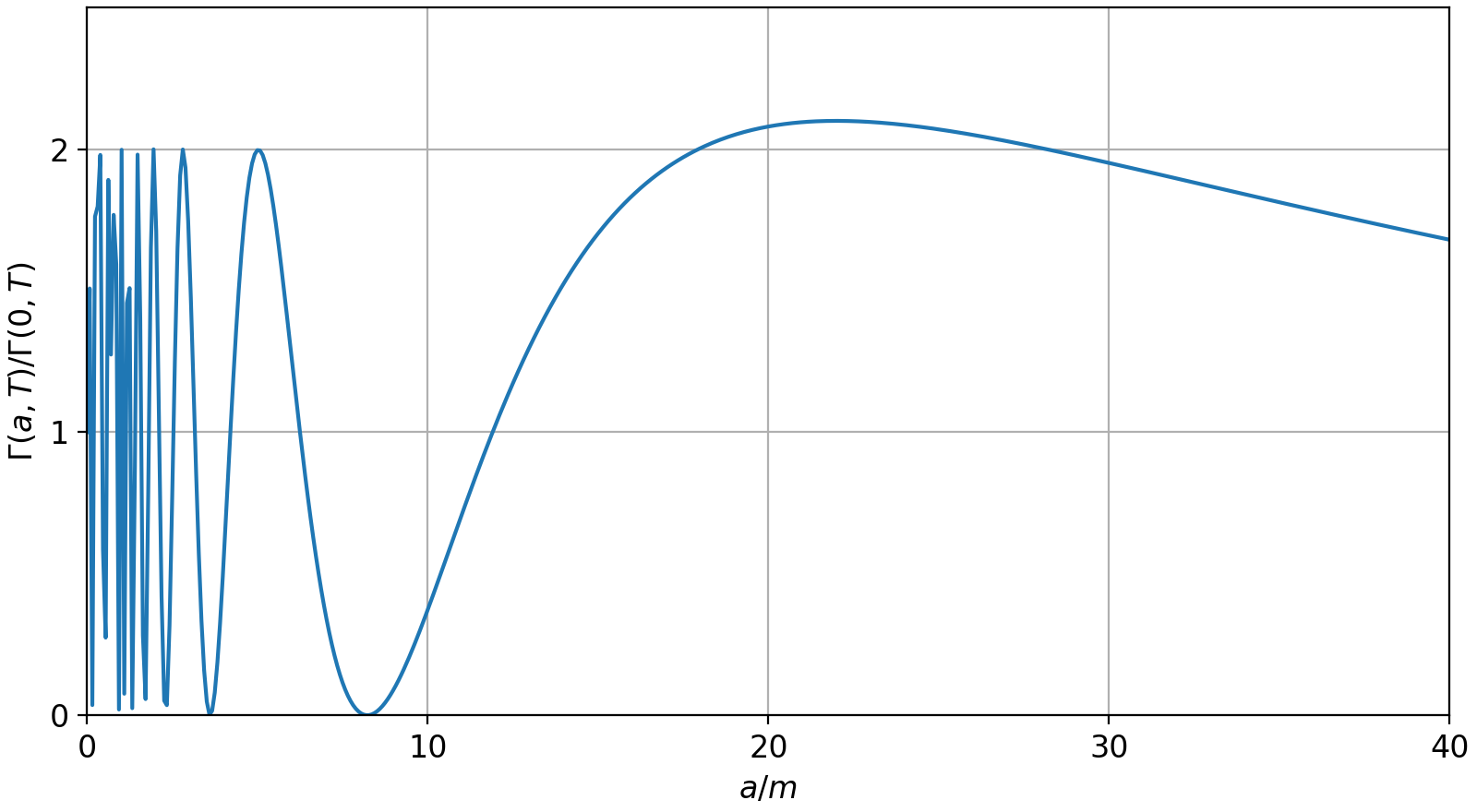}}
\caption{Infinite-time limit of the decay rate of an accelerated detector as function of acceleration, for a detector with $\frac{\Delta}{m}=10$. Note the peak above 2 in this case. Again, the rapid oscillations as $a$ goes to $0$ are not fully captured due to sampling limits.}
\label{fig:accelerated_infa2}
\end{figure}

Figure~\ref{fig:accelerated_a} shows clearly the peaks passing the detector's energy $\frac{\Delta}{m}$ as acceleration increases, finally showing a continuous dropoff once the first peak (leftmost in Figure~\ref{fig:accelerated_delta}) has passed. Furthermore, it also shows the importance in the accelerated case of the finite integration time, as this causes the low acceleration behaviour to properly converge to the non-accelerated limit. This is further confirmed by the plot of the $T\rightarrow\infty$ limit in Figures~\ref{fig:accelerated_infdelta} and~\ref{fig:accelerated_infa}. The nonphysical behaviour of the infinite-time limit clearly shows a need to use the finite-time formalism in this situation.

Note that, as also demonstrated in Figure~\ref{fig:accelerated_infa2}, whilst in the limit $a\rightarrow 0$ the oscillations are between $0$ and $2$ times the non-accelerated decay rate, this does not hold for the first peak(s).

A final observation to note is that, for constant detector energy gap, as acceleration gets large enough, the decay rate goes to zero. This happens because the first peak moves further and further right, becoming lower in the process, whilst at the same time the detector's measurement energy moves further away from the peak along the left flank.

\subsection{Scaling result}\label{subsec:scaling}

In the above analysis, we have five parameters introducing scales into the physics. For four of these parameters, $T$, $a$, $m$, $\Delta$, numerical stability in our plotting process limits us to values relatively close to unity. This will prove a limitation once considering physical realisations. To aleviate this somewhat, it is useful to note that under the transformation
\begin{align}
\bar{T} &= \frac{T}{C},\nonumber\\
\bar{a} &= aC,\nonumber\\
\bar{m} &= mC,\nonumber\\
\bar{\Delta} &= \Delta C,\nonumber\\
\bar{G} &= G C,
\end{align}
where $C$ is an arbitrary constant, we find
\begin{align}
\frac{\bar{P}_{10}(\bar{T})}{\bar{G}^2\left|\bra{0}M(0)\ket{1}\right|^2\bar{T}} &= \frac{1}{2\bar{T}}\int_0^{\bar{T}}\sd\kappa \left(\bar{T}-\kappa\right)\sin\left(\bar{\Delta}\kappa\right)J_0\left(\frac{2\bar{m}}{\bar{a}}\sinh\left(\frac{\bar{a}\kappa}{2}\right)\right)\nonumber\\
&\phantom{=}-\frac{1}{2\bar{T}}\int_0^{\bar{T}}\sd\kappa\left(\bar{T}-\kappa\right)\cos\left(\bar{\Delta}\kappa\right)Y_0\left(\frac{2\bar{m}}{\bar{a}}\sinh\left(\frac{\bar{a}\kappa}{2}\right)\right)\\
&= \frac{1}{2CT}\int_0^T\sd\kappa \left(T-\kappa\right)\sin\left(\Delta\kappa\right)J_0\left(\frac{2m}{a}\sinh\left(\frac{a\kappa}{2}\right)\right)\nonumber\\
&\phantom{=}-\frac{1}{2CT}\int_0^T\sd\kappa\left(T-\kappa\right)\cos\left(\Delta\kappa\right)Y_0\left(\frac{2m}{a}\sinh\left(\frac{a\kappa}{2}\right)\right)\\
&=\frac{1}{C}\frac{P_{10}(T)}{G^2\left|\bra{0}M(0)\ket{1}\right|^2T}
\end{align}

This result will allow us to rescale the parameters $T$, $a$, $m$, $\Delta$ such that they are closer to unity, absorbing the scaling factor into the coupling constant $G$. It is especially useful when considering decay rates relative to a non-accelerated baseline, as that causes the coupling constant to drop out.

\subsection{Patterns in the peaks}\label{sec:patterns}

\begin{figure}[htp]
\center{\includegraphics[width=0.9\textwidth]{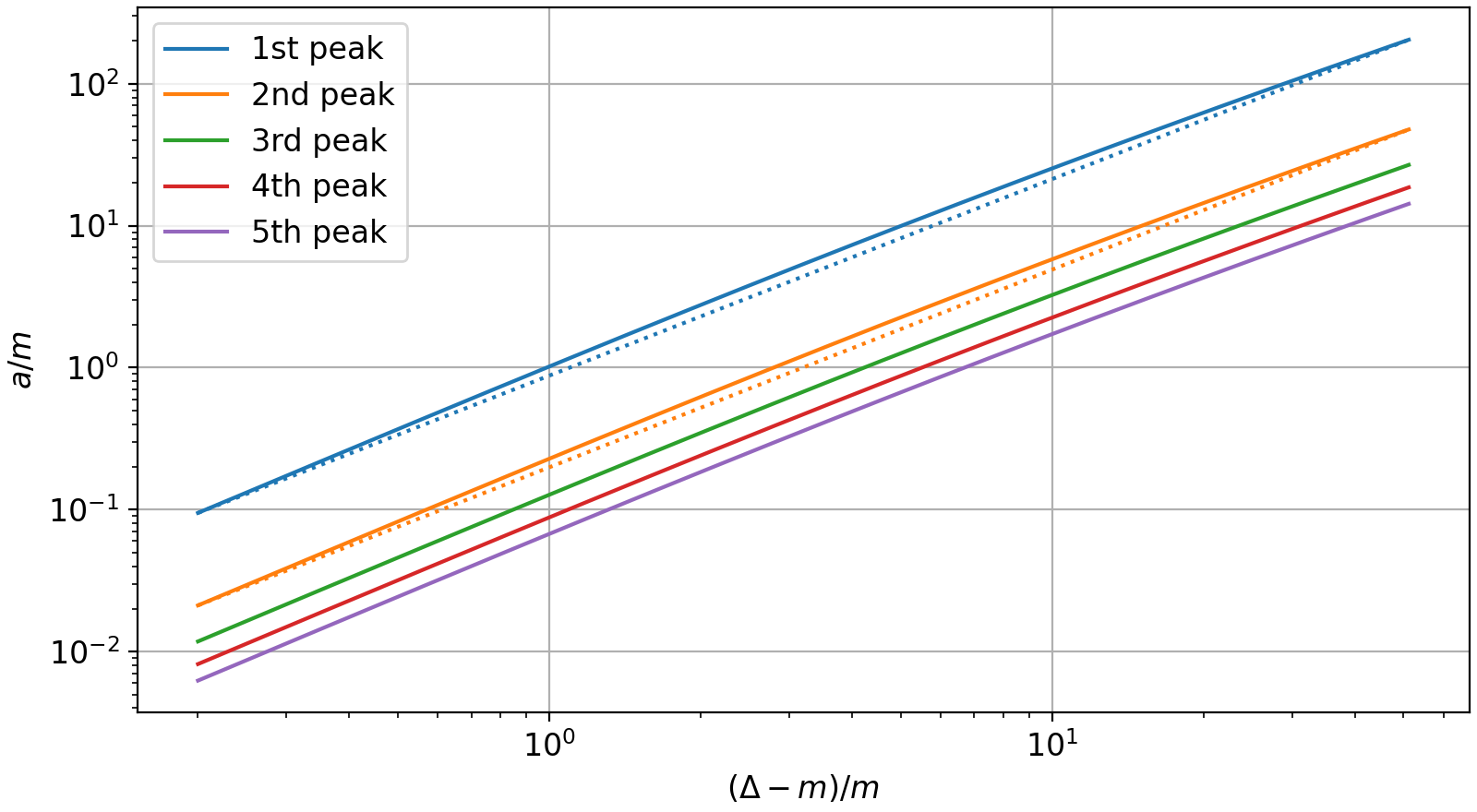}}
\caption{Location of the 5 peaks at highest acceleration as function of $\Delta-m$, at $T\rightarrow\infty$. For reference, a dotted straight line is drawn between the ends of the curves for the 1st and 2nd peaks.}
\label{fig:peaks_delta}
\end{figure}
\begin{figure}[htp]
\center{\includegraphics[width=0.9\textwidth]{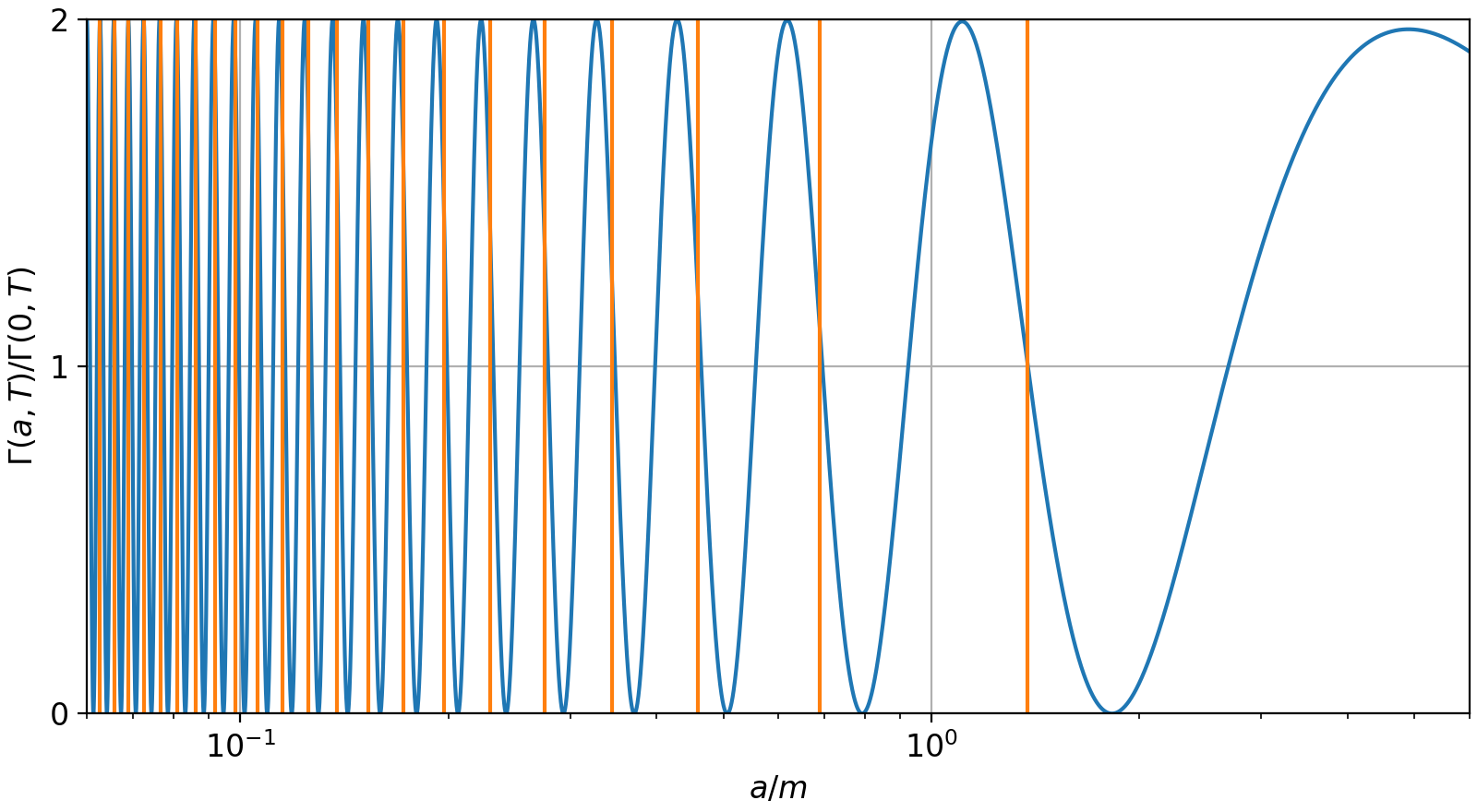}}
\caption{Locations of peaks predicted with $\frac{C}{n-1}$ pattern as orange lines, where the 23rd peak is used to fit $C=1.378$. Plotted against decay rate as function of acceleration for $\frac{\Delta}{m}=4$ at $T\rightarrow\infty$.}
\label{fig:peaks_n}
\end{figure}
\begin{figure}[htp]
\center{\includegraphics[width=0.9\textwidth]{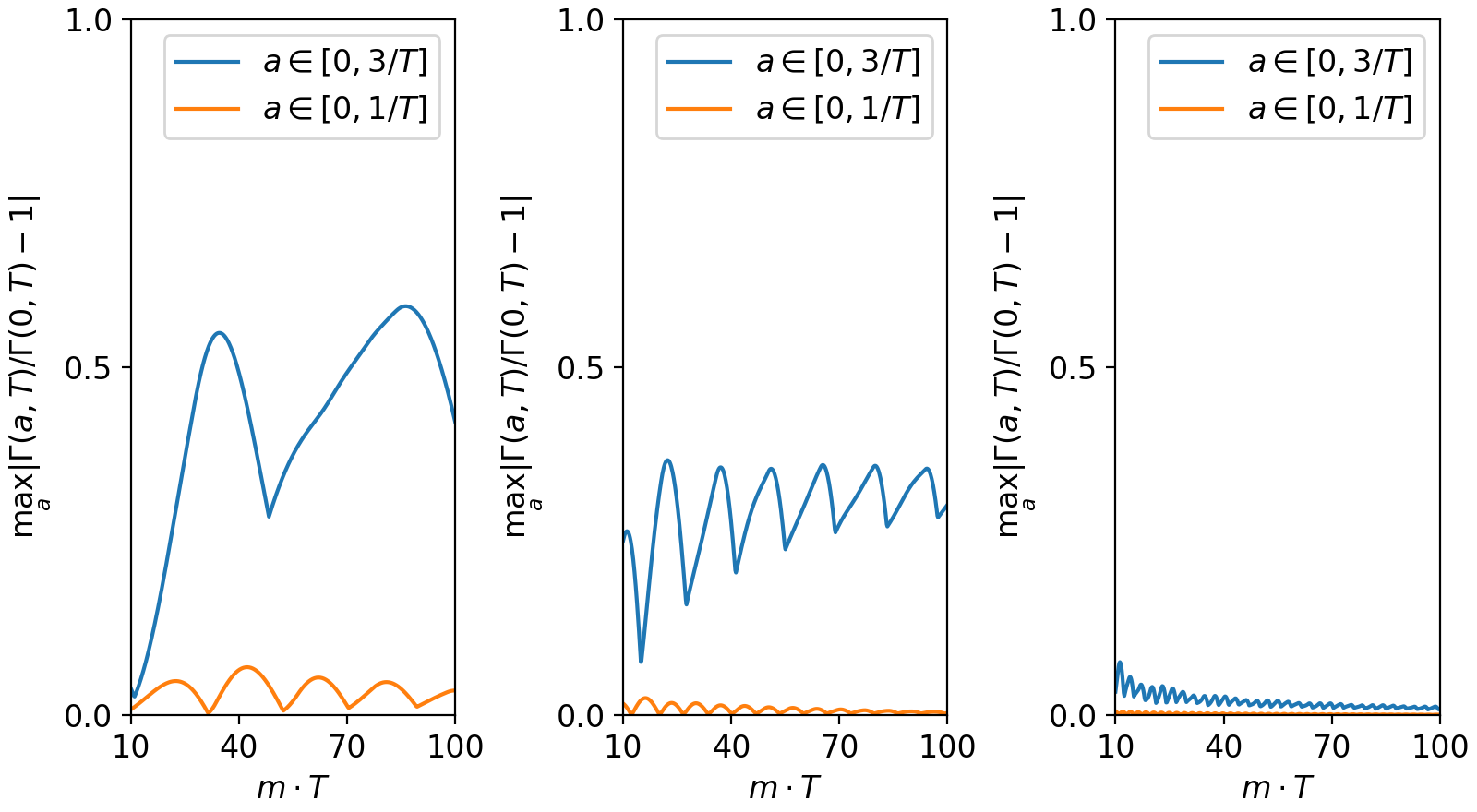}}
\caption{Maximum deviation $\left|\frac{\Gamma(a,T)}{\Gamma(0,T)}-1\right|$ of the decay rate observed in the indicated range of accelerations as a function of the amount of time the interaction is switched on. From left to right $\frac{\Delta}{m}=1.2$, $\frac{\Delta}{m}=1.5$ and $\frac{\Delta}{m}=2.5$.}
\label{fig:peaks_onset}
\end{figure}

As we will see in the next section, we will need to also be able to consider situations where both $\frac{\Delta-m}{m} << 1$ and $\frac{a}{m} << 1$. For $\frac{\Delta-m}{m} \lesssim 10^{-2}$ and $\frac{a}{m} \lesssim 10^{-2}$, this is out of reach of our current numerical methods, even when applying the above scaling result. Instead, we will focus here on three numerically observed patterns in the peaks of acceleration that will allow us to still make predictions in these cases. For this, we will number the peaks from right to left in the acceleration view, numbering the peak at the highest acceleration with $n=1$, and counting up from there.

First, as shown in Figure~\ref{fig:peaks_delta}, in the $T\rightarrow \infty$ limit the location of the peaks in acceleration space as a function of $\frac{\Delta-m}{m}$ follows a rough power law
\begin{align}
a_p(n, \frac{\Delta-m}{m}) \approx C(n) \times \left(\frac{\Delta-m}{m}\right)^e,
\end{align}
where $C(n)$ is a constant depending on which peak we look at. Fitting based on the location of the 5th peaks gives $e\approx 1.400$. This is not exact, and there is both some variation between the peaks (with the first peak giving an exponent closer to $1.388$, with a convergence to the value of the 5th peak for the other peaks) and deviation from the power law manifesting itself in a slightly convex\footnote{We intend the colloquial meaning of convex, not the mathematical definition for functions.} plot. Note that this bulging upward in the range plotted in Figure~\ref{fig:peaks_delta} implies that our estimate is likely to be an underestimate in that figure's range, and an overestimate outside it.

Second, as shown in Figure~\ref{fig:peaks_n}, in the $T\rightarrow \infty$ limit the locations of the peaks for fixed $\Delta-m$ can be approximated with $\frac{C}{n-1}$, where $C$ is a function of $\Delta-m$. Especially for higher $n$, this approximation becomes very accurate, and this will enable us to estimate the spacing between peaks.

Third, as can be seen in Figure~\ref{fig:accelerated_a}, for finite interaction durations, the low-acceleration peaks are supressed. In Figure~\ref{fig:peaks_onset}, we can see that this supression switches off in the region beyond $a=\frac{1}{T}$. This figure indicates that, for $\frac{\Delta-m}{m}<0.5$, we can expect to see at least a $20$ percent decay rate change when considering accelerations up to and including $a=\frac{3}{T}$.

The first two observations, together with the data on the 5th peak, can be combined to provide the following estimate of peak location at $T\rightarrow\infty$:
\begin{align}
\frac{a_p(n, \frac{\Delta-m}{m})}{m} \approx\frac{0.264 \left(\frac{\Delta-m}{m}\right)^{1.400}}{n-1}.
\end{align}
Within the range plotted in Figure~\ref{fig:peaks_delta}, this is accurate to about $15$ percent for the 5th peak, and to about $17$ percent for the 4th peak.

Together with the third observation, this will allow us to make qualitative predictions for physical realisations.

\section{Physical realisation}\label{sec:phys}

The model discussed in the previous section can be interpreted as a very rough approximation of alpha decay. This means approximating alpha particles as modes from a real scalar field with a single degree of freedom. The pre-decay nucleus is then taken to be the $\ket{1}$ state of the detector, and the post-decay nucleus is the $\ket{0}$ state of the detector. This is reasonable under the assumption that recoil from the decay does not significantly alter the trajectory of the particle.

Before looking at specific atoms, we reintroduce constants $\hbar$ and $c$ back into Formulas~\ref{eq:prob_nonaccel} and~\ref{eq:prob_accel}. This yields
\begin{align}
\frac{P_{10}(T)}{G^2\left|\bra{0}M(0)\ket{1}\right|^2T} &= \frac{1}{2T}\int_0^T\sd\kappa \left(T-\kappa\right)\sin\left(\frac{\Delta}{\hbar}\kappa\right)J_0\left(\frac{c^2m}{\hbar}\kappa\right)\nonumber\\
&\phantom{=}-\frac{1}{2T}\int_0^T\sd\kappa\left(T-\kappa\right)\cos\left(\frac{\Delta}{\hbar}\kappa\right)Y_0\left(\frac{c^2m}{\hbar}\kappa\right)
\end{align}
for a non-accelerated nucleus, and
\begin{align}
\frac{P_{10}(T)}{G^2\left|\bra{0}M(0)\ket{1}\right|^2T} &= \frac{1}{2T}\int_0^T\sd\kappa \left(T-\kappa\right)\sin\left(\frac{\Delta}{\hbar}\kappa\right)J_0\left(\frac{2c^3m}{\hbar a}\sinh\left(\frac{a\kappa}{2c}\right)\right)\nonumber\\
&\phantom{=}-\frac{1}{2T}\int_0^T\sd\kappa\left(T-\kappa\right)\cos\left(\frac{\Delta}{\hbar}\kappa\right)Y_0\left(\frac{2c^3m}{\hbar a}\sinh\left(\frac{a\kappa}{2c}\right)\right)
\end{align}
for an accelerated nucleus.

For alpha decay, $m$ will always be the mass of the alpha particle $3727.3794066\pm0.0000011\mathrm{MeV}/c^2$~\cite{codata}. We will further focus on the decay of $\mathrm{^{210}Po}$, which has a decay energy of $5.40753\pm 0.00007$~MeV~\cite{ame}. We choose to focus on this particular isotope as it is a pure alpha emitter decaying immediately to a stable isotope ($\mathrm{^{206}Pb}$), but our analysis can easily be generalized to other alpha emitters.

For this process, we thus have $\Delta = mc^2 + 5.407 MeV$, and hence $\frac{\Delta-c^2m}{c^2m} \approx 0.00145$, which puts us well outside the range of our numerical tools. However, we can still apply our observations from Section~\ref{sec:patterns}, which gives the following approximation for the locations of peaks when including constants:
\begin{align}
a_p\left(n, \frac{\Delta-c^2m}{c^2m}\right) &\approx \frac{c^3m}{\hbar}\frac{0.264\left(\frac{\Delta-c^2m}{c^2m}\right)^{1.400}}{n-1}.
\end{align}

This gives us rough approximations of where to expect peaks and, more importantly, through the distance between adjacent peaks we can also estimate the precision in acceleration needed to resolve individual peaks. Considering $\frac{a_p\left(n, \frac{\Delta-c^2m}{c^2m}\right) - a_p\left(n+1, \frac{\Delta-c^2m}{c^2m}\right)}{a_p\left(n, \frac{\Delta-c^2m}{c^2m}\right)}$ and simplifying, we find that the relative precision needed to resolve the $n$-th peak is roughly $\frac{1}{n}$.

Concretely, we find that the 100th peak is at an acceleration of approximately $4.8\times 10^{26} \frac{\mathrm{m}}{\mathrm{s}^2}$, and will require a precision of about 1 percent in the acceleration to resolve. Furthermore, based on observation 3 in Section~\ref{sec:patterns}, this peak will be visible with an integration time above $1.9$ attoseconds.

Although we look here at the 100th peak, higher-numbered peaks could be useful for experiments. However, this would come at the cost of requiring more precise acceleration to resolve those. Stated conversely, every order of magnitude we can increase the precision with which we accelerate the nuclei will enable us to resolve an order of magnitude more peaks, which in turn will enable a resolvable peak with about an order of magnitude lower acceleration.

\section{Conclusions and outlook}\label{sec:conclusion}

We have shown that, for a massive field in 1+1 dimensions, the de-excitation of an accelerated Unruh-deWitt detector shows significant non-trivial behaviour. Furthermore, when considering a de-exciting Unruh-deWitt detector as a model for alpha decay of a nucleus, we find that the decay-rate of the nucleus changes significantly with acceleration, varying between no decay at all and twice the normal decay rate or more at the peaks (Figures~\ref{fig:accelerated_delta} through~\ref{fig:accelerated_infa}).

Looking at the specific example of the alpha decay of $\mathrm{^{210}Po}$, we calculated that such effects could be measurable assuming the nucleus can be accelerated at a rate of about $4.8\times 10^{26} \frac{\mathrm{m}}{\mathrm{s}^2}$ with a relative precision of 1 percent. Comparing this to what the LHC is designed for, we find that it currently can achieve radial accelerations of about $2.9\times 10^{20}\frac{\mathrm{m}}{\mathrm{s}^2}$ for fully ionized $^{210}\mathrm{Po}$, with a precision of about $0.2$ percent~\cite{lhcdesign}.

Alternatives like spinning optical levitation experiments currently only reach accelerations of about $1.2*10^{14} \frac{\mathrm{m}}{\mathrm{s}^2}$~\cite{opticallev}. We will not go into plasma acceleration here as it currently seems mostly focussed around electron beams.

Although we still find 6 orders of magnitude between what can be achieved at the LHC and the effect we predict, there are multiple ways to significantly improve on this. First of all, although the LHC is currently the best we can do, proposals such as the FCC-hh\cite{fcchh} would improve about an order of magnitude in the achieved acceleration. Furthermore, colliders such as the LHC are optimized for high collision rates, rather than peak acceleration and precision in that peak acceleration. Each order of magnitude improvement in the relative precision of the achieved acceleration will result in approximately one order of magnitude lower acceleration required to see resolvable peaks.

A second avenue of improvement would be the energy scale of the decay process. The scaling law from Section~\ref{subsec:scaling} indicates that reducing the energy scale and mass of the decay process will reduce the required acceleration by a similar order of magnitude. This suggests it would be interesting to also consider beta decay processes, which would give a field mass of $511$KeV~\cite{codata} and, at the low end, decay energies around $15$KeV~\cite{ame}. Naively extrapolating our results to this, we would expect about 2 orders of magnitude reduction in required acceleration. However, beta decay has complications in the fact that it results in two particles, both of which are fermions. As such, expanding the results from this paper to situations with multiple fields and fermionic fields would be interesting directions for further research. The results in \cite{Matsas}, which show some modifications, are intruiging in this context, however the unorthodox fermion model makes it hard to estimate how applicable those results are.

Combining these two avenues suggest relatively straightforward ways to get 4 orders of magnitude gain over our results, leaving only a 3 orders of magnitude gap in acceleration still to be crossed with novel experimental approaches. This gives some hope that achieving observations of these effects may be possible in the future.

A caveat on the above is that our theoretical aproach is quite simplified. In this paper, we only studied the 1+1 dimensional case, and only looked at linear acceleration. We cannot exclude that these simplifications affect either the amplitude or pattern of the peaks we find. It would therefore be interesting in further studies to see how these effects change when considering the 3+1 dimensional case. Such an extension would also allow the study of circular motion, which for the regular Unruh effect has already been shown to give minor differences \cite{bell1}.

Beyond these open ends in connection with experimental realisations, our results also raise several theoretical questions. First of all, the observed pattern of interference peaks does not quite immediately provide a thermal interpretation. This raises the question how the deexcitation pattern we see here relates to the thermal spectra seen when considering excitations resulting from the Unruh effect.

Furthermore, given the parallels between Unruh radiation and Hawking radiation, it could be interesting to investigate whether there are similar parallels between de-excitation of accelerated detectors and detectors moving around in black hole geometries.

\bibliographystyle{unsrt}
\bibliography{accelerated_decay}

\begin{thebibliography}{10}

\bibitem{Fulling}
Stephen~A. Fulling.
\newblock {Nonuniqueness of Canonical Field Quantization in Riemannian
  Space-Time}.
\newblock {\em Phys. Rev. D}, 7:2850--2862, May 1973.

\bibitem{Unruh}
W.~G. Unruh.
\newblock {Notes on black-hole evaporation}.
\newblock {\em Phys. Rev. D}, 14:870--892, Aug 1976.

\bibitem{Crispino}
Lu\'{\i}s C.~B. Crispino, Atsushi Higuchi, and George E.~A. Matsas.
\newblock {The Unruh effect and its applications}.
\newblock {\em Rev. Mod. Phys.}, 80:787--838, Jul 2008.

\bibitem{bell1}
J.S. Bell and J.M. Leinaas.
\newblock {Electrons as accelerated thermometers}.
\newblock {\em Nuclear Physics B}, 212(1):131--150, 1983.

\bibitem{bell2}
J.S. Bell and J.M. Leinaas.
\newblock {The Unruh effect and quantum fluctuations of electrons in storage
  rings}.
\newblock {\em Nuclear Physics B}, 284:488--508, 1987.

\bibitem{Lynch1}
Morgan~H. Lynch, Eliahu Cohen, Yaron Hadad, and Ido Kaminer.
\newblock {Experimental observation of acceleration-induced thermality}.
\newblock {\em Phys. Rev. D}, 104:025015, Jul 2021.

\bibitem{Lynch2}
Morgan~H Lynch.
\newblock {Experimental observation of a Rindler horizon}.
\newblock {\em arXiv preprint arXiv:2303.14642}, 2023.

\bibitem{Zhou}
Yuebing Zhou, Jiawei Hu, and Hongwei Yu.
\newblock {Detecting circular Unruh effect with quantum entanglement}.
\newblock {\em arXiv preprint arXiv:2303.05638}, 2023.

\bibitem{Garay}
Luis~J. Garay, Eduardo Mart\'{\i}n-Mart\'{\i}nez, and Jos\'e de~Ram\'on.
\newblock {Thermalization of particle detectors: The Unruh effect and its
  reverse}.
\newblock {\em Phys. Rev. D}, 94:104048, Nov 2016.

\bibitem{Muller}
Rainer M\"uller.
\newblock {Decay of accelerated particles}.
\newblock {\em Phys. Rev. D}, 56:953--960, Jul 1997.

\bibitem{Matsas}
George E.~A. Matsas and Daniel A.~T. Vanzella.
\newblock {Decay of protons and neutrons induced by acceleration}.
\newblock {\em Phys. Rev. D}, 59:094004, Mar 1999.

\bibitem{DeWitt}
Bryce~S. deWitt.
\newblock {\em {Quantum Gravity: The New Synthesis}}, pages 680--745.
\newblock 1980.

\bibitem{DLMF}
{{\it NIST Digital Library of Mathematical Functions}}.
\newblock \url{https://dlmf.nist.gov/}, Release 1.1.9 of 2023-03-15.
\newblock F.~W.~J. Olver, A.~B. {Olde Daalhuis}, D.~W. Lozier, B.~I. Schneider,
  R.~F. Boisvert, C.~W. Clark, B.~R. Miller, B.~V. Saunders, H.~S. Cohl, and
  M.~A. McClain, eds.

\bibitem{codata}
Eite Tiesinga, Peter Mohr, David Newell, and Barry Taylor.
\newblock {CODATA Recommended Values of the Fundamental Physical Constants:
  2018}.
\newblock (93), June 2021.

\bibitem{ame}
Wang Meng, Audi G., F.~G. Kondev, Huang~W. J., S.~Naimi, and Xu~Xing.
\newblock {The AME2016 atomic mass evaluation (II). Tables, graphs and
  references}.
\newblock {\em Chinese Physics C}, 41(3):030003, March 2017.

\bibitem{lhcdesign}
Oliver~Sim Brüning, Paul Collier, P~Lebrun, Stephen Myers, Ranko Ostojic, John
  Poole, and Paul Proudlock.
\newblock {\em {{LHC Design Report}}}.
\newblock CERN Yellow Reports: Monographs. CERN, Geneva, 2004.

\bibitem{opticallev}
Yuanbin Jin, Jiangwei Yan, Shah~Jee Rahman, Jie Li, Xudong Yu, and Jing Zhang.
\newblock {6GHz hyperfast rotation of an optically levitated nanoparticle in
  vacuum}.
\newblock {\em Photon. Res.}, 9(7):1344--1350, July 2021.

\bibitem{fcchh}
FCC collaboration et~al.
\newblock {FCC-hh: the Hadron collider: future circular collider conceptual
  design report volume 3}.
\newblock {\em European Physical Journal: Special Topics}, 228(4):755--1107,
  2019.

\bibitem{GradRyzh}
I.~S. Gradshteyn and I.~M. Ryzhik.
\newblock {\em {Table of Integrals, Series, and Products}}.
\newblock Academic Press, 8th edition, 2014.

\end{thebibliography}

\appendix

\section{Infinite time limit}\label{app:inftimelimit}

\subsection{Accelerated detector}

To find the infinite time limit of the accelerated detector, first note that for large $x$, $\sinh(x) \propto e^x$, $J_0(x) \propto \frac{1}{\sqrt{x}}$ and $Y_0(x)\propto\frac{1}{\sqrt{x}}$. This, together with $Y_0(x)\propto\ln(x)$ for small x, implies
\begin{align}
\int_0^\infty \sd \kappa\;\kappa\sin\left(\Delta\kappa\right)J_0\left(\frac{2m}{a}\sinh\left(\frac{a\kappa}{2}\right)\right) < \infty,\\
\int_0^\infty \sd \kappa\;\kappa\cos\left(\Delta\kappa\right)Y_0\left(\frac{2m}{a}\sinh\left(\frac{a\kappa}{2}\right)\right) < \infty.
\end{align}

As a consequence, the $T\rightarrow\infty$ limit follows from the calculation
\begin{align}
\lim_{T\rightarrow\infty}\frac{P_{10}(T)}{G^2\left|\bra{0}M(0)\ket{1}\right|^2T} &= \lim_{T\rightarrow\infty} \frac{1}{2T}\int_0^T\sd\kappa\left(T-\kappa\right)\sin\left(\Delta\kappa\right)J_0\left(\frac{2m}{a}\sinh\left(\frac{a\kappa}{2}\right)\right)\nonumber\\
&\phantom{=\lim_{T\rightarrow\infty}}-\frac{1}{2T}\int_0^T\sd\kappa\left(T-\kappa\right)\cos\left(\Delta\kappa\right)Y_0\left(\frac{2m}{a}\sinh\left(\frac{a\kappa}{2}\right)\right)\\
&= \frac{1}{2}\int_0^\infty\sd\kappa\sin\left(\Delta\kappa\right)J_0\left(\frac{2m}{a}\sinh\left(\frac{a\kappa}{2}\right)\right)\nonumber\\
&\phantom{=}-\frac{1}{2}\int_0^\infty\sd\kappa\cos\left(\Delta\kappa\right)Y_0\left(\frac{2m}{a}\sinh\left(\frac{a\kappa}{2}\right)\right)\\
&=\frac{1}{a\pi}\sinh\left(\frac{\pi\Delta}{a}\right)\left(K_{\frac{i\Delta}{a}}\left(\frac{m}{a}\right)\right)^2\nonumber\\
&\phantom{=}+\frac{1}{a\pi}\cosh\left(\frac{\pi\Delta}{a}\right)\left(K_{\frac{i\Delta}{a}}\left(\frac{m}{a}\right)\right)^2\\
&=\frac{1}{a\pi}e^{\frac{\pi\Delta}{a}}\left(K_{\frac{i\Delta}{a}}\left(\frac{m}{a}\right)\right)^2,
\end{align}
where the second to last equality follows from equations 6.679.4 and 6.679.6 in \cite{GradRyzh}.

\subsection{Stationary detector}

The stationary case is slightly more subtle, as the integrals
\begin{align}
\int_0^\infty \sd x\;x\sin\left(ax\right)J_0\left(x\right),\\
\int_0^\infty \sd x\;x\cos\left(ax\right)Y_0\left(x\right)
\end{align}
do not converge.

To work around this, note that (\cite{DLMF} Section 10.6)
\begin{align}
\left(xJ_0\left(x\right)\right)' &= J_0\left(x\right)-xJ_1\left(x\right)\\
\left(xJ_1\left(x\right)\right)' &=xJ_0\left(x\right)\\
\left(xY_0\left(x\right)\right)' &= Y_0\left(x\right)-xY_1\left(x\right)\\
\left(xY_1\left(x\right)\right)' &=xY_0\left(x\right)
\end{align}
and hence, we can partially integrate twice to find
\begin{align}
\int xJ_0\left(x\right)\sin\left(ax\right)\sd x &= -\frac{1}{a}\cos\left(ax\right)xJ_0\left(x\right)-\frac{1}{a}\int xJ_1\left(x\right)\cos\left(ax\right)\sd x\nonumber\\
&\phantom{=}+\frac{1}{a}\int J_0\left(x\right)\cos\left(ax\right)\sd x\\
&= -\frac{1}{a}\cos\left(ax\right)xJ_0\left(x\right)-\frac{1}{a^2}\sin\left(ax\right)xJ_1\left(x\right)\nonumber\\
&\phantom{=}+\frac{1}{a^2}\int xJ_0\left(x\right)\sin\left(ax\right)\sd x+\frac{1}{a}\int J_0\left(x\right)\cos\left(ax\right)\sd x\\
&=-\frac{a}{a^2-1}\cos\left(ax\right)xJ_0\left(x\right)-\frac{1}{a^2-1}\sin\left(ax\right)xJ_1\left(x\right)\nonumber\\
&\phantom{=}+\frac{a}{a^2-1}\int J_0\left(x\right)\cos\left(ax\right)\sd x\\
\int xY_0\left(x\right)\cos\left(ax\right)\sd x &= \frac{1}{a}\sin\left(ax\right)xY_0\left(x\right)+\frac{1}{a}\int xY_1\left(x\right)\sin\left(ax\right)\sd x\nonumber\\
&\phantom{=}-\frac{1}{a}\int Y_0\left(x\right)\sin\left(ax\right)\sd x\\
&=\frac{1}{a}\sin\left(ax\right)xY_0\left(x\right)-\frac{1}{a^2}\cos\left(ax\right)xY_1\left(x\right)\nonumber\\
&\phantom{=}+\frac{1}{a^2}\int xY_0\left(x\right)\cos\left(ax\right)\sd x-\frac{1}{a}\int Y_0\left(x\right)\sin\left(ax\right)\sd x\\
&=\frac{a}{a^2-1}\sin\left(ax\right)xY_0\left(x\right)-\frac{1}{a^2-1}\cos\left(ax\right)xY_1\left(x\right)\nonumber\\
&\phantom{=}-\frac{a}{a^2-1}\int Y_0\left(x\right)\sin\left(ax\right)\sd x
\end{align}
where the last step for both follows from moving all factors proportional to the original integral to the left hand side, then dividing by the resulting proportionality factor. Note that the remaining integral on the right hand side is finite for all combinations of bounds on the positive real axis including 0.

This now allows us to calculate two expressions for limits. Starting with $J_0$:
\begin{align}
\lim_{T\rightarrow\infty} \frac{1}{T}\int_0^TxJ_0\left(x\right)\sin\left(ax\right)\sd x &= -\frac{a}{a^2-1}\lim_{T\rightarrow\infty}\cos\left(aT\right)J_0\left(T\right)\nonumber\\
&\phantom{=}-\frac{1}{a^2-1}\lim_{T\rightarrow\infty}\sin\left(aT\right)J_1\left(T\right)\nonumber\\
&\phantom{=}+\frac{a}{a^2-1}\lim_{T\rightarrow\infty}\frac{1}{T}\int_0^T J_0\left(x\right)\cos\left(ax\right)\sd x\\
&= 0.
\end{align}

The $Y_0$ case is slightly more complicated
\begin{align}
\lim_{T\rightarrow\infty} \frac{1}{T}\int_0^TxY_0\left(x\right)\cos\left(ax\right)\sd x &= \frac{a}{a^2-1}\lim_{T\rightarrow\infty}\sin\left(aT\right)Y_0\left(T\right)\nonumber\\
&\phantom{=}-\frac{a}{a^2-1}\lim_{T\rightarrow\infty}\frac{1}{T}\lim_{x\rightarrow 0}\sin\left(ax\right)xY_0\left(x\right)\nonumber\\
&\phantom{=}-\frac{1}{a^2-1}\lim_{T\rightarrow\infty}\cos\left(aT\right)Y_1\left(T\right)\nonumber\\
&\phantom{=}+\frac{1}{a^2-1}\lim_{T\rightarrow\infty}\frac{1}{T}\lim_{x\rightarrow 0}\cos\left(ax\right)xY_1\left(x\right)\nonumber\\
&\phantom{=}-\frac{a}{a^2-1}\lim_{T\rightarrow\infty}\frac{1}{T}\int_0^T Y_0\left(x\right)\sin\left(ax\right)\sd x\\
&= -\frac{1}{a^2-1}\lim_{T\rightarrow\infty}\frac{2}{\pi T}\nonumber\\
&= 0
\end{align}

This allows us to again conclude
\begin{align}
\lim_{T\rightarrow\infty}\frac{1}{2T}\int_0^T\sd\kappa\;\kappa\sin\left(\Delta\kappa\right)J_0\left(m\kappa\right) = 0\\
\lim_{T\rightarrow\infty}\frac{1}{2T}\int_0^T\sd\kappa\;\kappa\cos\left(\Delta\kappa\right)Y_0\left(m\kappa\right) = 0
\end{align}
and like before, we calculate
\begin{align}
\lim_{T\rightarrow\infty} \frac{P_{10}(T)}{G^2\left|\bra{0}M(0)\ket{1}\right|^2T} &= \lim_{T\rightarrow\infty}\frac{1}{2T}\int_0^T\sd\kappa\left(T-\kappa\right)\sin\left(\Delta\kappa\right)J_0\left(m\kappa\right)\nonumber\\
&\phantom{=\lim_{T\rightarrow\infty}}-\frac{1}{2T}\int_0^T\sd\kappa\left(T-\kappa\right)\cos\left(\Delta\kappa\right)Y_0\left(m\kappa\right)\\
&=\frac{1}{2}\int_0^\infty\sd\kappa\sin\left(\Delta\kappa\right)J_0\left(m\kappa\right)\nonumber\\
&\phantom{=}-\frac{1}{2}\int_0^\infty\sd\kappa\cos\left(\Delta\kappa\right)Y_0\left(m\kappa\right)\\
&=\begin{cases}
0 & \text{if } m > \Delta\\
\frac{1}{\sqrt{\Delta^2-m^2}} & \text{if } m < \Delta
\end{cases}
\end{align}
where the last step follows from equations 6.671.7 and 6.671.12 in \cite{GradRyzh}
\end{document}